\documentclass[prl,aps,twocolumn]{revtex4-1} 
\usepackage{graphicx} 
\usepackage{tabularx}
\usepackage{dcolumn} 
\usepackage{color}
\usepackage{bm}
\usepackage{amsmath}
\usepackage{amssymb}

\newcommand{\bi}[1]{\ensuremath{\boldsymbol{#1}}} 
\newcommand{\nn}{\nonumber}

\begin{document} 
 
\title{%
Topological spin texture and d-vector rotation in spin-triplet superconductors: A case of UTe$_2$}

\author{Yasumasa Tsutsumi} 
\affiliation{Department of Physics, Kwansei Gakuin University, Sanda, Hyogo 669-1330, Japan.} 
\author{Kazushige Machida} 
\affiliation{Department of Physics, Ritsumeikan University, Kusatsu 525-8577, Japan} 

\date{\today}

\begin{abstract}
A novel spin texture formed by Cooper pair spins is found theoretically with a phase string attached by half-quantized vortices at both ends
in a unit cell and characterized by its topologically rich vortex structure in a spin-triplet pairing.
It is stable at an intermediate field region sandwiched by two conventional singular vortex phases below and above it.
The d-vector direction of this spin texture is tilted from the principal crystal axes, whose spin susceptibility is neither the normal Pauli one $\chi_{\rm N}$ nor zero, describing microscopically the process of the d-vector rotation phenomena observed recently in UTe$_2$.
We compare the spin texture and singular vortex state in relation to the quasi-particle structure 
with Majorana zero modes for STM,
the nuclear spin resonance spectral line width for NMR and $\mu$SR, and the vortex form factors  for SANS to
facilitate the identification of the pairing symmetry in UTe$_2$.
\end{abstract}

\maketitle 


A spin-triplet pairing is extremely rare in nature. However, its utility is highly envisioned in fundamental physics to applications such as quantum computation, except for superfluid $^3$He~\cite{volovik,3he,mizushima}. 
 Only a few phases have been identified among the rich 3$\times$3$\times$2=18 dimensional order parameter space spanned by SO(3)$_{\rm spin}$$\times$SO(3)$_{\rm orbital}$$\times$U(1)$_{\rm gauge}$ 
such as ABM and BW, or more recently, their distorted versions in addition to the polar-~\cite{polar} and $\beta$-phases~\cite{beta}.
In this context, the pairing symmetry in a prime candidate of a spin-triplet pairing of UTe$_2$
 as a solid-state counterpart is quite intriguing, and its identification is critical.

The concept of the d-vector rotation is well established in neutral $^3$He superfluids~\cite{volovik,3he,mizushima}. 
The d-vector, which fully characterizes a spin-triplet pairing, changes its direction uniformly in bulk
under a magnetic field to minimize the Zeeman energy loss due to the Pauli paramagnetic effect acting on the Cooper pair spin.
Namely, the d-vector tries to remain perpendicular to the field orientation because the anisotropic spin susceptibility 
parallel to $\vec d$ is smaller than that perpendicular to $\vec d$.
In contrast, the situation is more complicated and physically richer for a spin-triplet superconductor with charged fermions because the applied field inevitably induces the depairing current acting on the Cooper pair orbital motion associated with vortices in addition to the Pauli effect.
Thus,  the d-vector rotation is realized by a vortex morphological change, or a spin textural change of Cooper pairs in general.
Although experimental observations have been reported recently in UTe$_2$~\cite{ishida1,ishida2,ishida3,ishida4,ishida5,ishida6} and UPt$_3$~\cite{tou1,tou2} 
a long time ago, where the Knight shift gradually changes as varying a field, its microscopic description has not been done before.
The experimental observations are interpreted only phenomenologically~\cite{nishira}.
This task is crucial in identifying the pairing symmetry and topological nature of materials.

UTe$_2$ is a newly found heavy fermion superconductor~\cite{ran} and is regarded as a spin-triplet pairing candidate.
Currently, there is much attention focussed on this strongly correlated system because of the following reasons:
(1) The Knight shift (KS), or the spin susceptibility $\chi_{\rm s}(T)$ decreases for all principal field directions, $a$, $b$, and $c$-axes at lower $H$~\cite{ishida1,ishida2,ishida3,ishida4,ishida5,ishida6}.
As a function of $H$, $\chi_{\rm s}(H)$ becomes gradually unchanged for $H\parallel b$- and $c$-axes at lower $T$. 
Thus, $\chi_{\rm s}(H)=\chi_{\rm N}$
with $\chi_{\rm N}$ normal spin susceptibility.
This is incompatible with the  $^3$He-B phase-like state because it simply means  the loss of the condensation energy 
by eliminating the component parallel to the field direction. However, this is physically improbable.
 Moreover, the magnitude of
the KS change for the $a$-axis is abnormally larger (four times the normal susceptibility) than 
the other $b$ and $c$ directions~\cite{ishida6}, implying that KS for the $H$$\parallel$$a$-axis is governed by a mechanism different 
from the ordinary KS drop, rather analogous to the Takagi mechanism observed in $^3$He-A phase~\cite{takagi}.
(2) The $H$-$T$ phase diagram for the $H$$\parallel$$b$-axis, which is the main focus of the present study,
  consists of three phases~\cite{georg,sakai,tokiwa,rosuel}:
low (LSC), intermediate (MSC), and high (HSC) field phases. 
They meet at a tetracritical point: $H_{\rm tetra}\sim$15 T.
$H_{\rm c2}$  for HSC has a prominent positive slope $(dT_{\rm c}/dH)_{H_{\rm c2}}>0$ emanating from $H_{\rm tetra}$.
KS remains unchanged (decreases) for HSC (LSC).
MSC in 15T$\lesssim$$H$$\lesssim$22T is characterized by a finite flux flow state where vortices are ``highly mobile''~\cite{sakai,tokiwa},
suggesting that an exotic coreless vortex is formed~\cite{tokiwa}.
Interestingly, it approximately corresponds to a field region in which $\chi_{\rm s}(H)$ is varying toward
$\chi_{\rm N}$ and reaching at $H\sim$15 T with increasing $H$.
Under pressure, the phase diagram has more than three distinct phases~\cite{daniel,pressure1}.
(3) There exists a variety of non-trivial superconducting properties.
The polar Kerr experiment shows that SC breaks time-reversal symmetry~\cite{kaptulnik}. The STM measurements detect the chiral symmetry-breaking state at sample edges~\cite{madhavan1} and pair-density wave~\cite{madhavan2}.
Several thermodynamic experiments suggest a nodal gap structure~\cite{review,metz,kittaka,shibauchi,matsuda}.
$H_{\rm c2}$ for three crystal directions far exceeds the Pauli limiting field~\cite{review}.
 
These experimental findings provide strong evidence for a spin-triplet pairing in UTe$_2$.
The order parameter must be properly configured to describe the d-vector rotation phenomenon microscopically.
Since $\chi_{\rm s}(T)$ decreases below T$_{\rm c}$ for three principal directions at lower fields, the d-vector must have three components with complex numbers whose transition temperatures associated with each component are nearly degenerate;
otherwise, continuous d-vector rotation is impossible.
Upon increasing the $H$$\parallel$$b(c)$-axis, 
$\chi_{\rm s}(H)(<\chi_{\rm N})$ starts to gradually increase toward $\chi_{\rm N}$ at approximately
5T (1T), and finally reaches $\chi_{\rm N}$ at approximately $H_{\rm rot}^b$=15T ($H_{\rm rot}^c$=5T).
Since $H_{\rm rot}^i$ ($i=b,c$) depends on the field orientations, those are slightly different, reflecting that
the original spin space symmetry SO(3)$_{\rm spin}$ is weakly broken~\cite{machida0,machida1,machida2,machida3,machida4}.
We ignore this weak symmetry breaking to investigate the generic features of the stable spin
texture and establish the concept of the d-vector rotation in the followings.

The purpose of this study is to find a stable spin texture in a spin-triplet pairing under SO(3)$_{\rm spin}$ symmetry,
by solving the microscopic quasi-classical Eilenberger equation for a spin-triplet pairing~\cite{tsutsumi} with full three components
of the order parameter. This turns out  to be rich topological features with Majorana zero modes.
We also establish the concept of the d-vector rotation in connection with UTe$_2$.
In particular, the recently found MSC for $H\parallel b$ with ``mobile vortices'' is explained in terms of the
above spin texture consisting of coreless vortices.

We can obtain physical quantities from the quasi-classical Green's function 
defined in the 4 $\times$ 4 matrix in particle-hole and spin spaces,
\begin{align}
\widehat{g}(\bi{k},\bi{r},\omega_n) = -i\pi
\begin{pmatrix}
\hat{g}(\bi{k},\bi{r},\omega_n) & i\hat{f}(\bi{k},\bi{r},\omega_n) \\
-i\underline{\hat{f}}(\bi{k},\bi{r},\omega_n) & -\underline{\hat{g}}(\bi{k},\bi{r},\omega_n)
\end{pmatrix},
\end{align}
depending on the direction of the relative momentum of a Cooper pair $\bi{k}$, the center-of-mass coordinate of the Cooper pair $\bi{r}$,
and the Matsubara frequency $\omega_n=(2n+1)\pi k_B T$.
It satisfies the normalization condition $\widehat{g}^2=-\pi^2\widehat{1}$.
$\widehat{g}(\bi{k},\bi{r},\omega_n)$ is calculated using the Eilenberger equation~\cite{eilenberger,schopohl,serene,fogelstrom,sauls},
\begin{multline}
-i\hbar\bi{v}(\bi{k})\cdot\bi{\nabla }\widehat{g} \\
= \left[
\begin{pmatrix}
\hat{K}-\mu_{\rm B}{\bm B}({\bm r})\cdot\hat{\bm \sigma} & -\hat{\Delta }(\bi{k},\bi{r}) \\
\hat{\Delta }(\bi{k},\bi{r})^{\dagger } & -\hat{K}-\mu_{\rm B}{\bm B}({\bm r})\cdot\hat{\bm \sigma}
\end{pmatrix}
,\widehat{g} \right],
\label{Eilenberger eq}
\end{multline}
where 
$\hat{K}\equiv\hat{K}(\bi{k},\bi{r},\omega_n)=\left[i\omega_n+e\bi{v}(\bi{k})\cdot\bi{A}(\bi{r})\right]\hat{1}$.
%
%
It includes the paramagnetic effects due to the Zeeman term $\mu_{\rm B}{\bm B}({\bm r})\cdot\hat{\bm \sigma}$, where $\mu_{\rm B}$ is a renormalized Bohr magneton, ${\bm B}({\bm r})$ is the flux density of the internal field, and $\hat{\bm \sigma}$ is the Pauli matrix.
In this study, we consider the two-dimensional cylindrical Fermi surface, ${\bm k}=k_{\rm F}(\cos\theta,\sin\theta)$, with the Fermi velocity ${\bm v}({\bm k})=v_{\rm F}(\cos\theta,\sin\theta)$, where $0\le\theta<2\pi$.

We assume the chiral $p$-wave pairing symmetry with the pairing function $\phi({\bm k})=e^{i\theta}$ independent of the spin state.
In the assumption, the spin-triplet order parameter is given by
$\hat{\Delta }(\bi{k},\bi{r})=i\phi(\bi{k})\bi{d}(\bi{r})\cdot\hat{\bi{\sigma }}\hat{\sigma_y}$,
%
%
with the local $d$-vector ${\bm d}({\bm r})$.
The self-consistent condition for $\hat{\Delta }(\bi{k},\bi{r})$ is given as
$\hat{\Delta }(\bi{k},\bi{r}) = 
N_0\pi k_BT\sum_{|\omega_n| \le \omega_c}\left\langle V(\bi{k}, \bi{k}')\hat{f}(\bi{k}',\bi{r},\omega_n)\right\rangle_{\bi{k}'}
$
%
%
where $N_0$ is the DOS in the normal state,
$\omega_c$ is the cutoff energy setting $\omega_c=20\pi k_B T_c$ with the transition temperature $T_c$,
and $\langle\cdots\rangle_{\bi{k}}$ indicates the Fermi surface average.

The pairing interaction $V(\bi{k}, \bi{k}')=g\phi(\bi{k})\phi^*(\bi{k}')$, where $g$ is a coupling constant.
In our calculation, 
we use the relation $1/gN_0=\ln (T/T_c)+\pi k_BT\sum_{|\omega_n| \le \omega_c}1/|\omega_n|$.
%
%
The vector potential for the internal magnetic field $\bi{A}(\bi{r})$ is also self-consistently determined by
%
\begin{multline}
\nabla\times[\nabla\times\bi{A}(\bi{r})]={\bm \nabla}\times {\bm M}_{\rm para}({\bm r})\\
-i2eN_0\pi k_BT\sum_{|\omega_n|\le\omega_c}
\langle \bi{v}(\bi{k}) g_0(\bi{k},\bi{r}, \omega_n) \rangle_{\bi{k}},
\label{vector potential}
\end{multline}
where $g_0$ is a component of the quasi-classical Green's function $\hat{g}$ in spin space, namely,
\begin{align}
\hat{g} =
\begin{pmatrix}
g_0+g_z & g_x-ig_y \\
g_x+ig_y & g_0-g_z
\end{pmatrix}
=g_0\hat{1}+{\bm g}\cdot\hat{\bm \sigma}. \nn
\end{align}
The paramagnetic moment is given by 
$${\bm M}_{\rm para}({\bm r})={\bm M}_0({\bm r})+i2\mu_{\rm B}N_0\pi k_BT\sum_{|\omega_n|\le\omega_c}
\langle  {\bm g}(\bi{k},\bi{r}, \omega_n) \rangle_{\bi{k}}$$
%
%
where the normal state paramagnetic moment ${\bm M}_0({\bm r})=2\mu_{\rm B}^2N_0{\bm B}({\bm r})$.

We solve eq.~\eqref{Eilenberger eq}, the gap equation and ~\eqref{vector potential} alternately,
and obtain self-consistent solutions, under a given unit cell of the vortex lattice~\cite{nagato,schopohl2,ldos,ichioka2,ichioka3}
(see the detail in Supplements).
%
%
%
%
The local density of states (LDOS) for the energy $E$ is given by
$N_{\sigma}(\bi{r},E)=N_0 \left\langle {\rm Re} \left[g_{\sigma}^{\rm R}(\bi{k},\bi{r}, E)\right] \right\rangle_{\bi{k}}$,
%
%
with the retarded quasiclassical Green's function $\hat{g}^{\rm R}(\bi{k},\bi{r},E)\equiv\hat{g}(\bi{k},\bi{r}, \omega_n)|_{i\omega_n \rightarrow E+i\eta}$, where $\eta$ is a positive infinitesimal constant.
Here, $\sigma$ indicates the spin state and $g_{\sigma}^{\rm R}$ is defined as $g_{\uparrow}^{\rm R}\equiv g_0^{\rm R}+g_z^{\rm R}$ and $g_{\downarrow}^{\rm R}\equiv g_0^{\rm R}-g_z^{\rm R}$.
To obtain $g_{\sigma}^{\rm R}(\bi{k},\bi{r}, E)$, 
we solve eq.~\eqref{Eilenberger eq} with $E+i\eta$ instead of $i\omega_n$ under the pair potential and vector potential obtained self-consistently with $\eta=0.03E_0$.

\begin{figure}
\begin{center}
\includegraphics[width=8.5cm]{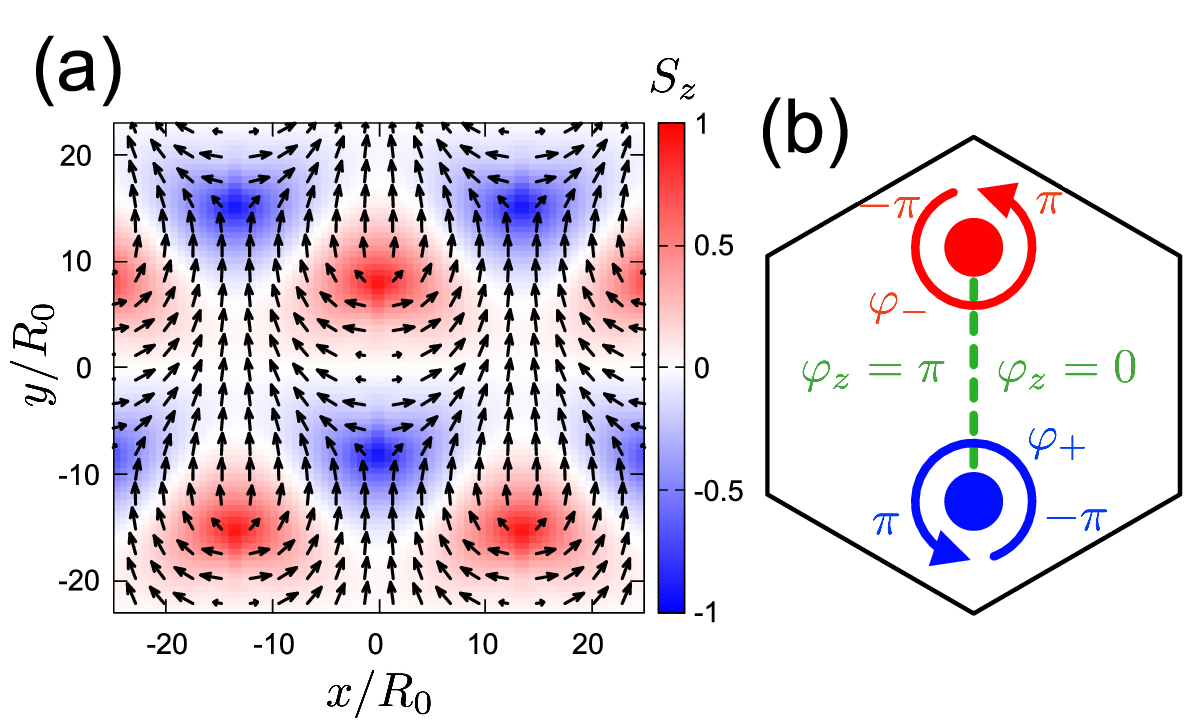}
\end{center}
\caption{\label{fig1}
(a) Spin  texture ${\bm S}=i({\bm d}\times{\bm d}^*)/|{\bm d}|^2$ with $\bar{B}=0.01B_0$, $\mu=1$, and $T=0.2T_{\rm c}$.
(b) Schematic picture for phases of d-vector components in a unit cell of the hexagonal vortex lattice.
The green dotted line indicates the phase string across which the phase of $\eta_z$ jumps by $\pi$.
Both ends, indicated by red and blue dots, correspond to the half-quantized vortices.
}
\end{figure}

Local spin polarization of the Cooper pair is given by the local d-vector as ${\bm S}({\bm r})=i[{\bm d}({\bm r})\times{\bm d}({\bm r})^*]/|{\bm d}({\bm r})|^2$.
We obtain a stable spin texture by the self-consistent calculation.
The spin texture under the magnetic field $\bar{B}=0.01B_0$ with $\mu=1$ at $T=0.2T_{\rm c}$ is shown in Fig.~\ref{fig1}.
The area of the unit cell of the hexagonal vortex lattice is $200\pi R_0^2$ under this magnetic field.
The spin texture has a position with $S_z=+1$ and a position with $S_z=-1$ in the unit cell.
The positions correspond to the phase singularity of d-vector components, $d_z=|d_z|e^{i\varphi_z}$ and $d_{\pm}=(\mp d_x+id_y)/\sqrt{2}=|d_{\pm}|e^{i\varphi_{\pm}}$.

A schematic picture for the phases, $\varphi_z$ and $\varphi_{\pm}$, in a unit cell is shown in Fig.~\ref{fig1}(b).
The singularities of $\varphi_+$ and $\varphi_-$ with $2\pi$-phase winding are situated at $(0,-y_{\rm c})$ and $(0,+y_{\rm c})$, respectively, with $y_{\rm c}\approx 5.6R_0$.
Phase $\varphi_z$ has singularities with $\pi$-phase winding also at $(0,-y_{\rm c})$, and $(0,+y_{\rm c})$ and phase jumps by $\pi$ at the branch cut on $x=0$ connecting these two singularities.
Thus, the phase of the d-vector changes by $2\pi$ around the unit cell of the vortex lattice. 

The phase singularities at $(0,\pm y_{\rm c})$ can be viewed as a half-quantized vortex (HQV)
because locally $d_-e^{i\varphi_-}+d_+=e^{i\varphi_-/2}\{d_x\cos(\varphi_-/2)+d_y\sin(\varphi_-/2)\}$.
Therefore, the spin texture obtained is described as the two HQVs (red and blue dots in Fig.~\ref{fig1}(b)) bound together by the string or branch cut (dotted green line) 
with the phase $\pi$ jump. The amplitude of the order parameters in the spin texture is non-vanishing anywhere.
As $H$ increases, the string length decreases, and the objects merge to form a singular vortex,
corresponding to the HSC phase where the spin vector is completely polarized along the field direction $z$.
We can calculate the average d-vector direction over the unit cell, finding the tilting angle 10.6$^{\circ}(11.3^{\circ})$ for $\bar{B}=0.01(0.03)B_0$ away from the field direction $z$.

The local spin of the Cooper pair is polarized in the $+y$-direction perpendicular to the field direction $z$
on the boundary edge of the hexagonal unit cell shown in Fig.~\ref{fig1}(b).
In contrast, the d-vector at $(0,0)$ is parallel to the $y$-axis with ${\bm S}={\bm 0}$.
The spin polarization is shrunk around $(0,0)$ with $|{\bm S}|<1$.
Note that $\Delta(\bm r)$ is nonvanishing everywhere, as displayed in Fig.~\ref{fig4}(a).
Thus, this spin texture is coreless with the soft cores, which is contrasted 
with the singular vortex states, having the hard core, stabilized in lower (LSC) and higher fields (HSC) ( see Supplements for detail).

\begin{figure}
\begin{center}
\includegraphics[width=8.5cm]{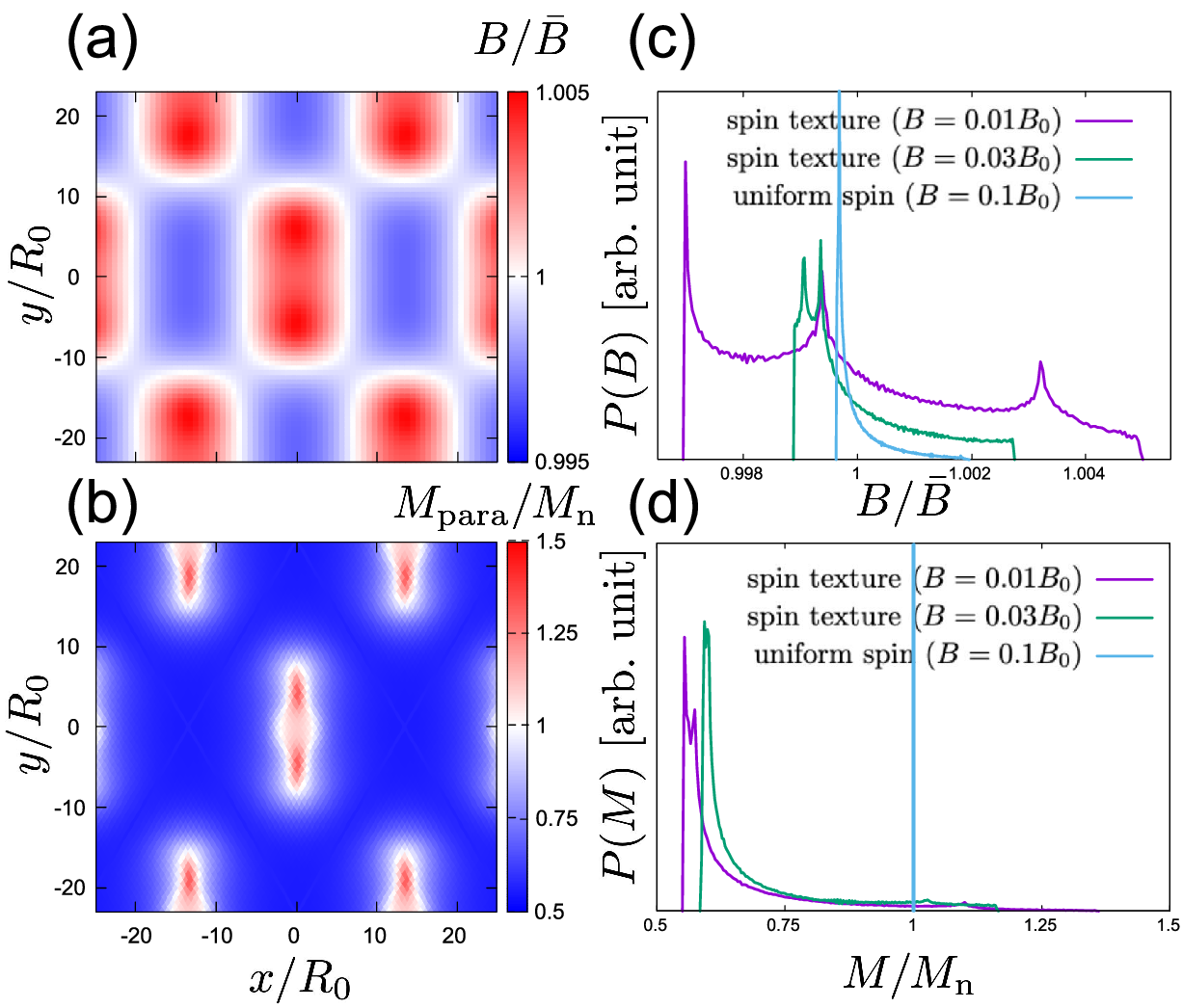}
\end{center}
\caption{\label{fig2}
Spatial variations of internal field $B({\bm r})$ (a) and paramagnetic moment $M_{\rm para}({\bm r})$ (b) with $\bar{B}=0.01B_0$, $\mu=1$, and $T=0.2T_{\rm c}$.
Distributions of internal field $P(B)$ (c) and paramagnetic moment $P(M_{\rm para})$(d) with $\mu=1$ and $T=0.2T_{\rm c}$.
The distributions are shown for spin textures under $\bar{B}=0.01B_0$ and $\bar{B}=0.03B_0$, and singular vortex structure with uniform spin parallel to the $z$-axis under $\bar{B}=0.1B_0$.
}
\end{figure}

Internal field $B({\bm r})$ and paramagnetic moment $M_{\rm para}({\bm r})$ for the spin texture are shown in Fig.~\ref{fig2}(a).
The internal field has maxima at the positions of the phase singularity; namely, the spin of the Cooper pair is polarized to the $z$-axis.
As shown in Fig.~\ref{fig2}(a), the central maximum region is thus elongated along the polarization direction of the $y$-axis.
The spatial dependence of the internal field gives the characteristic distribution function $P(B)\equiv\int\delta(B-B({\bm r}))d{\bm r}$, whose intensity originates from the length of a contour satisfying $B=B({\bm r})$.
$P(B)$ for the spin texture under $\bar{B}=0.01B_0$ has three peaks as shown in Fig.~\ref{fig2}(c).
The right peak around $B=1.003{\bar B}$ originates from a contour enclosing each maximum of the internal field.
The central peak slightly below $B=\bar{B}$ originates from the square contour colored with white in Fig.~\ref{fig2}(a).
The left peak at the lowest $B$ comes from the minima of the internal field.
In contrast, $P(B)$ for the singular vortex structure with uniform spin has a single peak due to internal field minima, as shown in Fig.~\ref{fig2}(c). 
Thus $P(B)$ for the spin texture is more widely distributed than that for the singular vortex structure.
The right peak at the highest $B$ in the distribution function for the spin texture disappears under $\bar{B}=0.03B_0$.
This is  because contours enclosing each maximum of the internal field become short owing to the maxima overlapping each other in the small area of the unit cell of the vortex lattice. 

These remarkable three peak structures in $P(B)$ can be observable by the muon spin rotation ($\mu$SR) experiment~\cite{sonier}.
This significantly impacts the SANS experiment, as fundamental form factors ~\cite{ichioka2} such as $F_{10}$ and $F_{20}$ increase as $H$ increases, in sharp contrast to conventional SC, where these decrease exponentially with $H$~\cite{ichioka2,ichioka3} because the two singularities approach each other. Thus,
 the elongated high field region in Fig.~\ref{fig2}(a)
becomes round to concentrate it, causing those form factors to grow.

The paramagnetic moment $M_{\rm para}({\bm r})$ shown in Fig.~\ref{fig2}(b)
also has  maxima at the positions of the phase singularity similar to $B({\bm r})$ mentioned above.
However, it is more strongly confined near the phase singularities, as shown from Fig.~\ref{fig2}(b).
$M_{\rm para}({\bm r})$ approaches $M_{\rm n}/2$ on the edge of the hexagonal unit cell 
because one of two components of the d-vector parallel to the $x$- and $z$-axes is parallel to the magnetic field.
The distribution functions of $M_{\rm para}({\bm r})$, $P(M)\equiv\int\delta(M-M_{\rm para}({\bm r}))d{\bm r}$, for the spin texture and the singular vortex structure are shown in Fig.~\ref{fig2}(d).
$P(M)$ for the spin texture has double peaks in low $M$ compared to  the singular vortex structure case.
The double peaks originates from long contours enclosing the minima of the paramagnetic moment
compared to the sharp single peak in the singular vortex.

$P(M)$ is directly measured by NMR on $^{125}$Te as KS~\cite{ishida2}.
The gradual change of KS with field for the $H\parallel b$-axis is interpreted as the d-vector rotation~\cite{ishida2,ishida3}.
Here, we have shown theoretically that this is indeed the case:
Starting at the lower field for LSC with  $\chi_{\rm s}$=$\chi_{\rm N}/2$ where all the d-vectors are parallel to $z$,
the peak position moves up from $M=M_{\rm n}/2$ with increasing $H$ as shown in Fig.~\ref{fig2}(d).
During this process, the d-vector rotates away from the field direction in MSC.
The average d-vector direction over a unit cell for the spin texture is tilted away from the field direction.
 The shift of the peak position of $P(M)$
precisely corresponds to the d-vector angle away from $\bf{B}$.
This process of the spin textural changes continues until all the local spin polarizations ${\bf S}({\bf r})$
direct along $\bf{B}$ with the peak at $M=M_{\rm n}$.
At higher fields $P(M)$ has a single sharp peak at the normal state position: $M=M_{\rm n}$ as shown in Fig.~\ref{fig2}(d),
corresponding to HSC, with all the d-vectors perpendicular to $z$.
The resulting vortex has a singularity at the vortex core, bearing the genuine Majorana zero mode,
as explained in Supplements. 
The one-to-one correspondence between the tilting angle and the peak position deviation from the bottom at $M_{\rm n}/2$
in KS establishes the concept of the d-vector rotation phenomenon.


\begin{figure}
\begin{center}
\includegraphics[width=8.5cm]{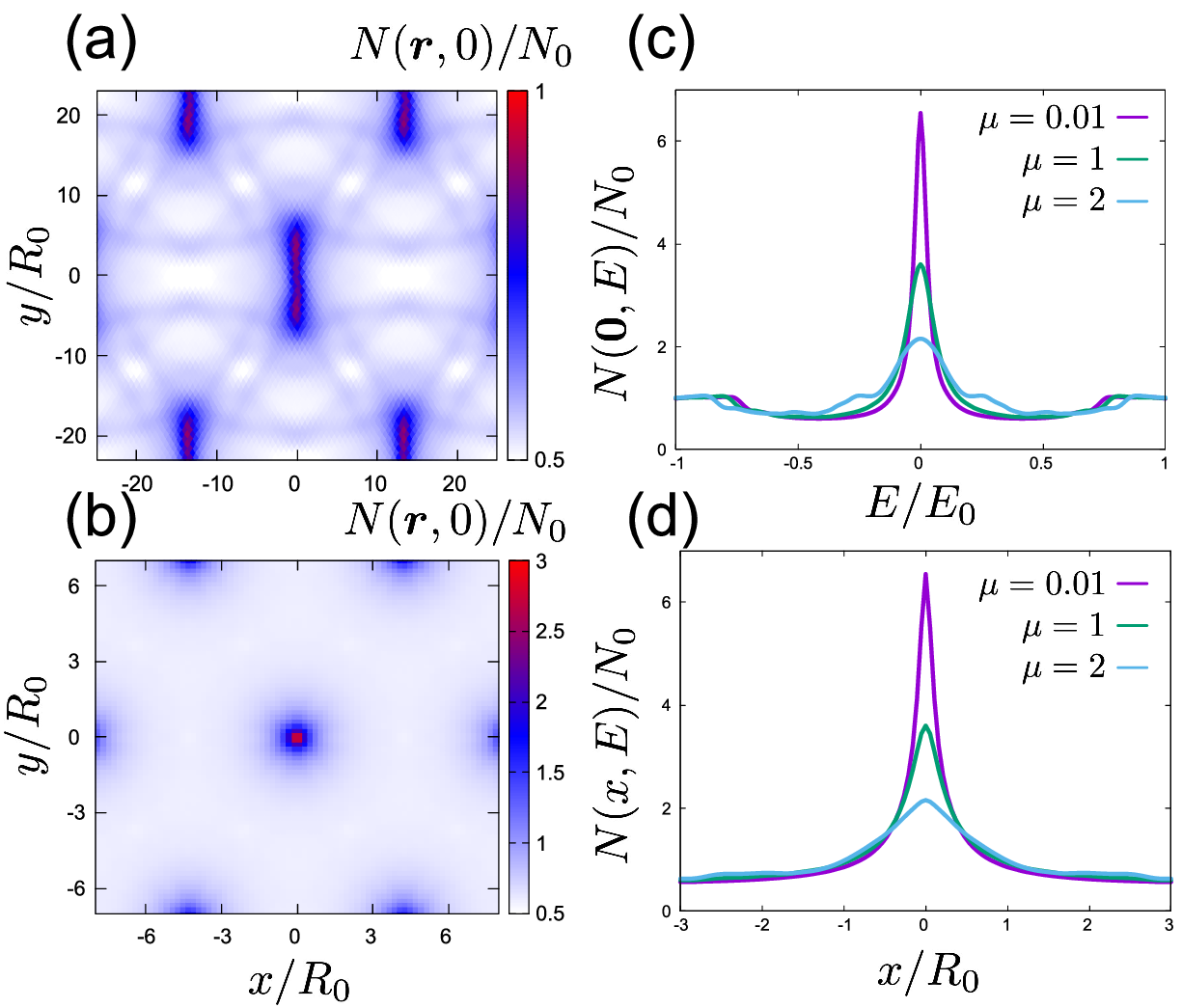}
\end{center}
\caption{\label{fig3}
Zero energy LDOS for the spin texture under $\bar{B}=0.01B_0$ (a) and singular vortex structure with uniform spin parallel to the $y$-axis under $\bar{B}=0.1B_0$ (b) with $\mu=1$ and $T=0.2T_{\rm c}$.
Energy profiles of the LDOS $N({\bm 0},E)$ at the vortex core (c) and  spatial variations of the zero energy LDOS $N(x,0)$ along the $x$-axis (d) for the singular vortex structure with $\mu=0.01$, $\mu=1$, and $\mu=2$.
}
\end{figure}

The zero energy LDOS for the spin texture with $\bar{B}=0.01B_0$, $\mu=1$, and $T=0.2T_{\rm c}$ is shown in Fig.~\ref{fig3}(a).
The zero energy quasiparticles are localized around the phase singularity of the d-vector, and along the branch cut, or phase-string
attached to the two singularities at both ends across which the phase of $\eta_z$ jumps by $\pi$.
This accords with the Jackiw-Rebbi theorem~\cite{jackiw} and its condensed matter version~\cite{nakanishi,fujita,mizu}.
In contrast, the zero energy quasiparticles are concentrated in the vortex core for the singular vortex structure.
The concentration of the zero energy LDOS is clearly shown in Fig.~\ref{fig3}(b) for the singular vortex structure with ${\bm S}\parallel{\bm y}$ under $\bar{B}=0.1B_0$ at $T=0.2T_{\rm c}$ with $\mu=1$.
The dependence of the LDOS on the strength of the Pauli paramagnetic effect, $\mu$, is found in the energy profile of the LDOS at the vortex core [Fig.~\ref{fig3}(c)] and the spatial variation of the zero energy LDOS along the $x$-axis [Fig.~\ref{fig3}(d)].
The intensity of the zero energy LDOS at the vortex core decreases as the Pauli paramagnetic effect increases.
The suppression of the zero energy peak height is accompanied by the broadening of the low energy spectrum, as shown in Fig.~\ref{fig3}(c).
A spatial spread of the zero energy LDOS around the vortex core is less sensitive to the strength of the Pauli paramagnetic effect [Fig.~\ref{fig3}(d)].
In contrast, the Pauli paramagnetic effect does not affect the LDOS for the singular vortex structure with ${\bm S}\parallel{\bm z}$.
Then, the LDOS for the singular vortex with ${\bm S}\parallel{\bm z}$ is equivalent to that with ${\bm S}\parallel{\bm y}$ for $\mu=0$.
Note that the LDOS is larger than $N_0/2$ because one of the spin states remains in the normal state in the spin-polarized superconductivity.

A singular vortex with ${\bm S}\parallel{\bm z}$ in HSC has a spin-polarized Majorana zero energy mode regardless of the Pauli paramagnetic effect.
The Pauli paramagnetic effect hybridizes a spin-polarized zero energy mode in the vortex core with quasiparticles in the normal spin state for a singular vortex with ${\bm S}\parallel{\bm y}$ in LSC.
Despite the hybridization of the spin states, their energy levels do not shift from zero energy.
The zero energy modes in a singular vortex with ${\bm S}\parallel{\bm y}$ are protected by the symmetry of the magnetic $\pi$-rotation around the $x$-axis.
The magnetic $\pi$-rotation symmetry gives a topological winding number that guarantees the existence of two zero energy modes in a singular vortex with ${\bm S}\parallel{\bm y}$ (see the detail in Supplements).

\begin{table}[t]
  \caption{Characterizations of the three phases}
  \label{table}
  \centering
  \begin{tabular}{lccccc}
 \hline  
phase & $\chi_s/\chi_N$ &d-vector&singularity&Majorana&width*\\
 \hline   \hline
 LSC & 1/2 &d$\parallel$$B$ (uniform)&yes&core&wide\\
MSC & 1/2$\sim$1 & spin texture&no&string&wide\\
HSC & 1 &d$\perp$$B$ (uniform)&yes&core&narrow\\
\hline
  \end{tabular}
*width: NMR resonance width 
\end{table}

In summary, we microscopically described the d-vector rotation phenomena
by solving the Eilenberger equation for a spin-triplet pairing  to understand this concept.
The obtained spin texture ${\bm S}(\bm r)$ for MSC stabilized in between the two singular vortex states with
uniform $\bm S$ for LSC (HSC) with $\bm S \perp \bm B$ ( $\bm S \parallel \bm B$) was fully characterized,
providing several important indications to understand experimental data on UTe$_2$. Table I summarizes
the characteristics of the three phases.
In particular, the newly found MSC experimentally characterized by an unusual flux flow state, or  mobile vortices~\cite{sakai,tokiwa},
might correspond to this spin texture, consisting of two coreless HQV's tagged by the phase string.
We demonstrated the rich topological features, including the locations of the Majorana zero modes
 associated with the spin texture to be investigated experimentally in UTe$_2$.

One (KM) of the authors thanks K. Ishida, S. Kitagawa, H. Sakai, D. Aoki, Y. Shimizu, S. Kittaka, T.
Sakakibara, Y. Tokunaga, Y. Haga, and A. Miyake for helpful discussions.
In particular, K.  Ishida allowed us to share their unpublished data which motivated the present project.
This work was supported by JSPS KAKENHI, No.17K05553 and No. 21K03455, and
 performed in part at Aspen Center for Physics, which is supported by National Science Foundation grant PHY-2210452.

\clearpage
\setcounter{section}{0}
\setcounter{figure}{0}
\setcounter{equation}{0}
\renewcommand{\thefigure}{S\arabic{figure}}
\renewcommand{\theequation}{S.\arabic{equation}}
\renewcommand{\thetable}{S\arabic{table}}
\renewcommand{\thesection}{S\arabic{section}}

\section{Supplements}

\subsection{I.  Self-consistent calculations and detailed numerics}


The unit cell is divided into $41\times 41$ mesh points,
where we obtain the quasi-classical Green's functions, ${\bm d}(\bi{r})$, $\bi{A}(\bi{r})$, and ${\bm M}_{\rm para}({\bm r})$.
When we solve eq.~\eqref{Eilenberger eq} by the Riccati method~\cite{nagato,schopohl2},
we estimate ${\bm d}(\bi{r})$, and $\bi{A}(\bi{r})$ at arbitrary positions by the interpolation from their values at the mesh points,
and by the periodic boundary condition of the unit cell including the phase factor due to the magnetic field~\cite{ldos,ichioka2,ichioka3}.
In the numerical calculation, we use the units $R_0=\hbar v_F/(2\pi k_BT_c)$, $B_0=\hbar/(2|e|R_0^2)$, and $E_0=\pi k_BT_c$
for the length, magnetic field, and energy, respectively. 
By the dimensionless expression, eq.~\eqref{vector potential} is rewritten as
\begin{multline}
\frac{R_0}{B_0}\nabla\times[\nabla\times\bi{A}(\bi{r})]=\frac{R_0}{B_0}\nabla\times\bi{M}_{\rm para}(\bi{r})\\
+i\frac{1}{\kappa^2}\frac{T}{T_c}\sum_{|\omega_n|\le\omega_c}
\left\langle \frac{\bi{k}}{k_{\rm F}} g_0(\bi{k},\bi{r}, \omega_n) \right\rangle_{\bi{k}},
\end{multline}
with
\begin{align}
\bi{M}_{\rm para}(\bi{r})=\bi{M}_0(\bi{r})+i\frac{\mu B_0}{\kappa^2}\frac{T}{T_c}\sum_{|\omega_n|\le\omega_c}
\left\langle  {\bm g}(\bi{k},\bi{r}, \omega_n) \right\rangle_{\bi{k}},
\end{align}
where $\mu=\mu_{\rm B}B_0/E_0$ and $\kappa=B_0/(E_0\sqrt{2 N_0})=\sqrt{7\zeta(3)/8}\kappa_{\rm GL}$. 
We use a large GL parameter $\kappa_{\rm GL}=60$.

Since magnetic fields are applied to the $z$-direction, in the symmetric gauge, the vector potential $\bi{A}(\bi{r})=(\bar{\bi{B}}\times\bi{r})/2+\bi{a}(\bi{r})$,
where $\bar{\bi{B}}=(0,0,\bar{B})$ is a uniform flux density and $\bi{a}(\bi{r})$ is related to the internal field $\bi{B}(\bi{r})=\bar{\bi{B}}+\nabla\times\bi{a}(\bi{r})$.
We consider the hexagonal vortex lattice whose coordinate in the unit cell is given by

\begin{align}
\bi{r}=s_1(\bi{u}_1-\bi{u}_2)+s_2\bi{u}_2
\end{align}
%


\noindent
with $-0.5\le s_i\le 0.5$ $(i=1,2)$,
$\bi{u}_1=\ell(1,0,0)$, and $\bi{u}_2=\ell(1/2,\sqrt{3}/2,0)$.

The spin texture is obtained by the self-consistent calculation from the following initial spatial structure of the d-vector.
The initial d-vector has the singularities of $\varphi_z$, $\varphi_+$, and $\varphi_-$ at $(0,0)$, $(0,-3R_0)$, and $(0,+3R_0)$, respectively, with $2\pi$-phase winding.
We constrain the d-vector to remain in the stable spin-polarized state except within a radius of $3R_0$ around $(0,0)$.
The free region is introduced to converge in a smooth structure.
The singularities of $\varphi_{\pm}$ allow to move to the favorable position, and the singularity of $\varphi_z$ splits into two singularities with $\pi$-phase winding in the converged process.
Note that a singular vortex structure with a uniformly polarized spin state is a possible self-consistent solution in this calculation.
The singular vortex structure is stabilized under high magnetic field or at high temperature.
As a result of the large length scale of the spin-polarized rotation region, the spin texture is unfavorable for a small unit cell of the vortex lattice.

\subsection{II. Singular vortex}

\begin{figure}
\includegraphics[width=8.5cm]{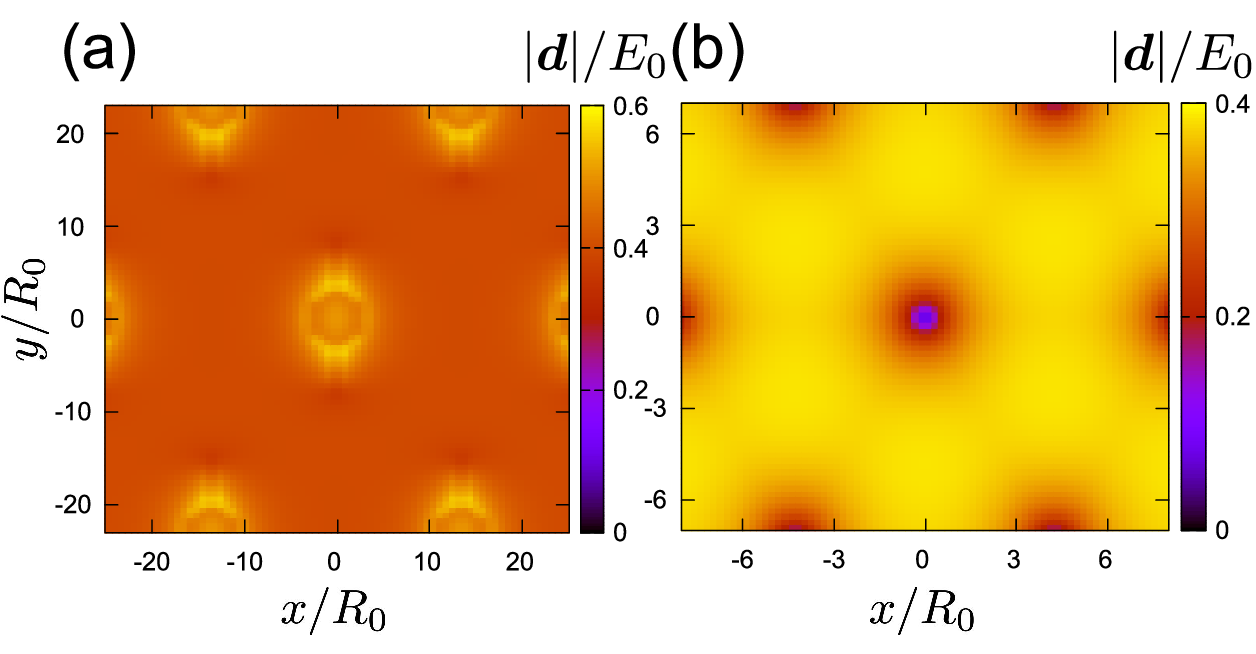}
\caption{\label{fig4}
Spatial profiles of the order parameter $\Delta({\bf r})/\pi k_{\rm B}T_{\rm c}$ for the spin texture, $\bar{B}=0.01B_0$, $\mu =1.0$ and $T=0.2T_{\rm c}$ (a),
and for the singular vortices in the uniform spin state, $\bar{B}=0.1B_0$, $\mu =1.0$ and $T=0.2T_{\rm c}$ (b).
}
\end{figure}

\begin{figure}
\includegraphics[width=8.5cm]{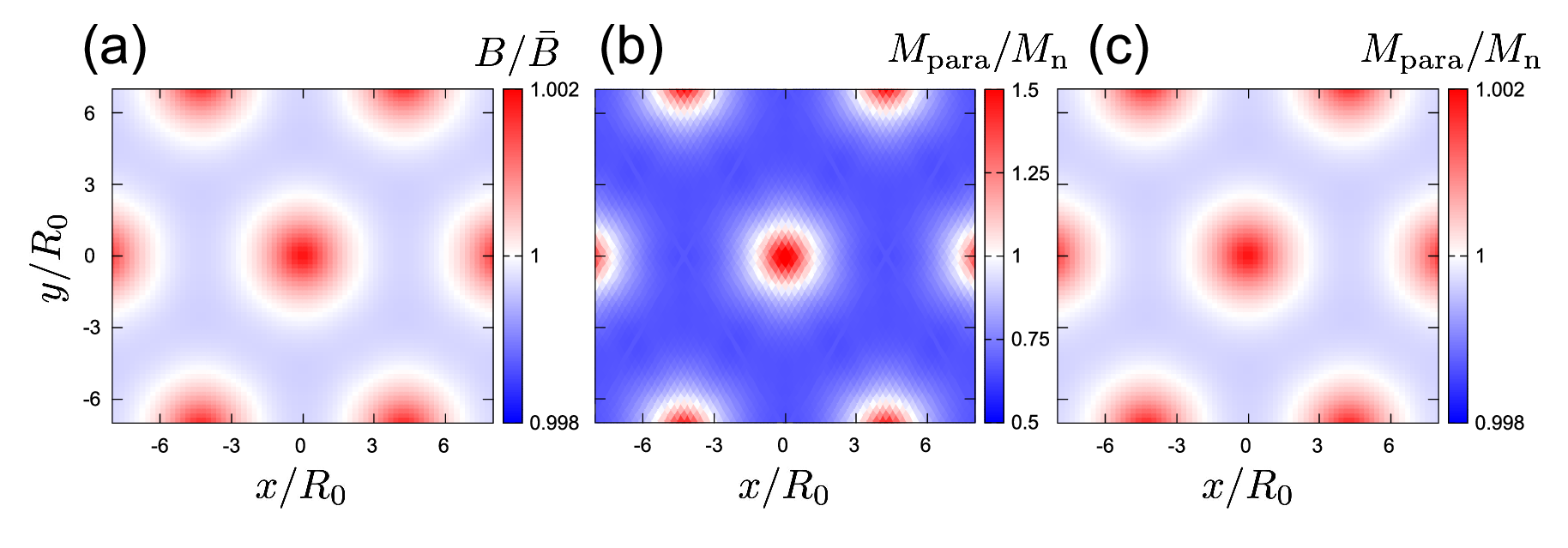}
\caption{\label{fig5}
Internal field distribution $B({\bf r})/\bar{B}$ for the singular vortices with ${\bf s}\perp {\bf B}$
and ${\bf s}\parallel {\bf B}$ in the uniform spin states (a), $M_{\rm para}({\bf r})/M_{\rm n}$ for ${\bf s}\perp {\bf B}$ (b),
and $M_{\rm para}({\bf r})/M_{\rm n}$ for ${\bf s}\parallel {\bf B}$ (c).
}
\end{figure}

As shown in Fig.~\ref{fig4}(a), the order parameter $\Delta({\bf r})$ 
is non-vanishing everywhere in the obtained spin texture, showing that it is a soft core, or coreless vortex state.
This is because the three components compensate each other
to gain the condensation energy maximally.
In contrast, in the vortex states for ${\bf S}\perp{\bf B}$ and ${\bf S}\parallel{\bf B}$ shown in Fig.~\ref{fig4}(b),
$\Delta({\bf r})$ vanishes at the vortex center; those are hard core vortex states.
This singular vortex state is more susceptible to vortex pinning at impurities or defects than that
for the spin texture.
The Majorana zero mode is localized in the vortex core  both for ${\bf S}\perp{\bf B}$ and ${\bf S}\parallel{\bf B}$.
As shown in Fig.~\ref{fig5}(a), the internal distribution $B({\bf r})$ is identical 
for ${\bf S}\perp{\bf B}$ and ${\bf S}\parallel{\bf B}$, while $M_{\rm para}({\bf r})$ is different for the two cases, as shown in Figs.~\ref{fig5}(b) and (c). This causes different $P(M)$ profiles, leading to the different Knight shift values,
namely, $\chi_{\rm s}\simeq\chi_{\rm n}/2$ for the former and $\chi_{\rm s}=\chi_{\rm n}$ for the latter.


\subsection{III. Topological classification of zero energy modes in a singular vortex}

Consider a vortex along the $z$-axis, and a circle surrounding the vortex, which is parametrized by the angle $\phi$ evaluated from the $x$-axis.
The semiclassical Bogoliubov-de Gennes (BdG) Hamiltonian on the circle is given by
\begin{align}
\widehat{H}_{\rm BdG}({\bm k},\phi)=
\begin{pmatrix}
\hat{h}({\bm k}) & \hat{\Delta}({\bm k},\phi) \\
\hat{\Delta}^{\dagger}({\bm k},\phi) & -\hat{h}^{\rm T}(-{\bm k})
\end{pmatrix},
\end{align}
where $\hat{h}({\bm k})=(\hbar^2/2m)({\bm k}^2-k_{\rm F}^2)\hat{1}-\mu_{\rm B}B\hat{\sigma}_z$ is the Hamiltonian in the normal state, and $\hat{\Delta}$ is OP which approaches
\begin{align}
\hat{\Delta}({\bm k},\phi)=i\Delta_0\frac{k_x+ik_y}{k_{\rm F}}\frac{1}{\sqrt{2}}(\hat{\sigma}_z+i\hat{\sigma}_x)\cdot\hat{\sigma}_ye^{i\phi},
\end{align}
far away from the vortex core.
The BdG Hamiltonian has particle-hole symmetry defined by $\widehat{\mathcal{C}}\widehat{H}_{\rm BdG}({\bm k},\phi)\widehat{\mathcal{C}}^{-1}=-\widehat{H}_{\rm BdG}^*(-{\bm k},\phi)$.
The singular vortex has the magnetic $\pi$-rotation symmetry around the $x$-axis, which implies $\widehat{\mathcal{R}}\widehat{H}_{\rm BdG}({\bm k},\phi)\widehat{\mathcal{R}}^{-1}=\widehat{H}_{\rm BdG}^*(-k_x,k_y,-\phi)$.
From the combination of the magnetic $\pi$-rotation symmetry and the particle-hole symmetry, the BdG Hamiltonian obeys  $(\widehat{\mathcal{C}}\widehat{\mathcal{R}})\widehat{H}_{\rm BdG}({\bm k},\phi)(\widehat{\mathcal{C}}\widehat{\mathcal{R}})^{-1}=-\widehat{H}_{\rm BdG}(k_x,-k_y,-\phi)$.
The combined symmetry defined the chiral symmetry $\widehat{\Gamma}\widehat{H}_{\rm BdG}(k_x,k_y=0),\phi=0,\pi)\widehat{\Gamma}^{-1}=-\widehat{H}_{\rm BdG}(k_x,k_y=0),\phi=0,\pi)$ with $\widehat{\Gamma}=\widehat{\mathcal{C}}\widehat{\mathcal{R}}$ in the symmetric space $k_y=0$ and $\phi=0$ or $\pi$.
The chiral symmetry enables us to introduce the one-dimensional winding number as follows:
\begin{multline}
w^{\phi=0,\pi}=-\frac{1}{4\pi i}\int dk_x{\rm tr}\left[\widehat{\Gamma}\widehat{H}_{\rm BdG}^{-1}(k_x,k_y=0),\phi=0,\pi)\right.\\
\left.\times\partial_{k_x}\widehat{H}_{\rm BdG}(k_x,k_y=0),\phi=0,\pi)\right].
\end{multline}
For the singular vortex with ${\bm S}\parallel{\bm y}$, the one-dimensional winding number is evaluated as $w^0=2$ and $w^{\pi}=-2$ under the Pauli paramagnetic effect, $\mu_{\rm B}B\ne 0$.
The difference of the winding $(w^0-w^{\pi})/2=2$ provides the number of zero energy modes in the singular vortex core.
For the singular vortex without Pauli paramagnetic effect, $\mu_{\rm B}B=0$, the difference of the winding $(w^0-w^{\pi})/2=1$ implies that a spin-polarized Majorana zero energy mode is localized in the vortex core.
Note that we can evaluate the one-dimensional winding number for a singular vortex with ${\bm S}\parallel{\bm z}$, in which the difference of the winding $(w^0-w^{\pi})/2=1$, regardless of the Pauli paramagnetic effect.

\end{document}